\documentclass[aps,pre,reprint,longbibliography,floatfix,groupedaddress]{revtex4-1}

\usepackage[pdftex]{graphicx}
\usepackage{hyperref}
\usepackage{amsmath,amssymb,amsfonts}

\begin{document}

\title{Percolation thresholds for discorectangles: numerical estimation for a range of aspect ratios}

\author{Yuri~Yu.~Tarasevich}
\email[Corresponding author: ]{tarasevich@asu.edu.ru}

\author{Andrei~V.~Eserkepov}
\email{dantealigjery49@gmail.com}

\affiliation{Laboratory of Mathematical Modeling, Astrakhan State University, Astrakhan, Russia, 414056}

\date{\today}

\begin{abstract}
Using Monte Carlo simulation, we have studied the percolation of discorectangles. Also known as  stadiums or two-dimensional spherocylinders, a discorectangle is a rectangle with semicircles at a pair of opposite sides. Scaling analysis was performed to obtain the percolation thresholds in the thermodynamic limits. We found:  (i) for the two marginal aspect ratios $\varepsilon = 1$ (disc) and $\varepsilon \to \infty$ (stick) the percolation thresholds coincide with known values within the statistical error; (ii) for intermediate values of $\varepsilon$ the percolation threshold lies between the percolation thresholds for ellipses and rectangles and approaches the latter as the aspect ratio increases.
\end{abstract}

\maketitle

\section{Introduction}\label{sec:intro}
Percolation, i.e., the emergence of a connected subset (a cluster) that spans opposite boundaries in a disordered medium, has attracted the attention of the scientific community for several decades~\cite{Stauffer,Sahimi1994,BollobasRiordan2006,Grimmett1999,Kesten1982}. The occurrence of a percolation cluster drastically changes the physical properties of the medium, e.g., an insulator--conductor phase transition can be observed when the disordered medium is a mixture of conductive and insulating substances. Special attention has been paid to percolation in disordered systems produced by the random deposition of elongated particles onto a substrate~\cite{Li2009PRE,Mertens2012PRE,Li2013PRE,Li2016PhysA}. Elongated species such as nanotubes, nanowires, and nanorods are of particular interest for nanotechnology, e.g., the production of transparent electrodes~\cite{Hecht2011AM,Nam2016N,Ackermann2016SR,Callaghan2016PCCP,McCoul2016AEM,Zhang2017JMM,Hicks2018JAP}.

To characterize a deposit, the number density, i.e., the number of objects, $N$, per unit area, $A$, is commonly used
\begin{equation}\label{eq:numdensity}
  n = \frac{N}{A}.
\end{equation}
Another useful quantity is the filling fraction,
\begin{equation}\label{eq:fillingfraction}
  \eta = n a,
\end{equation}
where $a$ is the area of one particle.
The total fraction of the plane covered by the overlapping (penetrable) particles is
\begin{equation}\label{eq:coveredfraction}
  \phi = 1 - \exp(-\eta)
\end{equation}
(see, e.g.,~\cite{Mertens2012PRE}).

\begin{figure}[!htb]
  \centering
  \includegraphics[width=0.5\columnwidth]{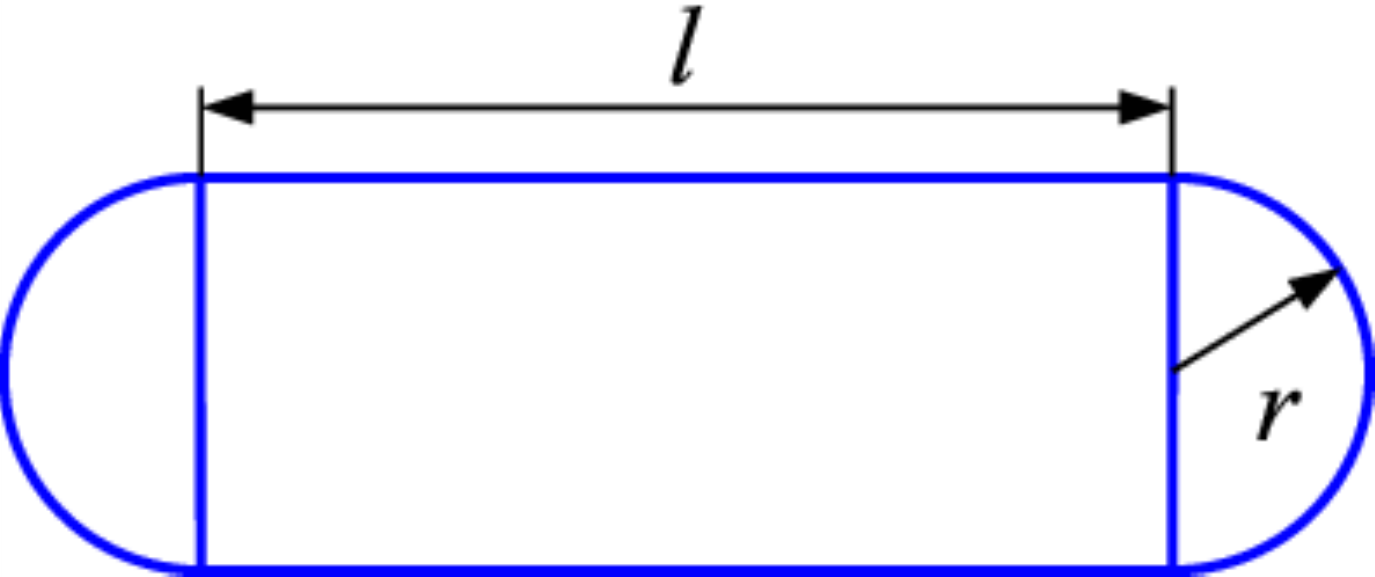}
  \caption{Example of a discorectangle.\label{fig:stadium}}
\end{figure}

To mimic the shape of elongated particles and, at the same time, simplify the simulations, different simple geometrical figures are used, e.g., sticks, rectangles, ellipses, superellipses, and discorectangles. A discorectangle is a rectangle with semicircles at a pair of opposite sides (figure~\ref{fig:stadium}). Its aspect ratio is
\begin{equation}\label{eq:asperctratio}
  \varepsilon = 1 + \frac{l}{2r}.
\end{equation}
A discorectangle (or ``stadium'') is a two-dimensional analog of a spherocylinder (a ``stadium of revolution'' or ``capsule''), i.e., a three-dimensional geometric shape consisting of a cylinder with hemispherical ends.

Percolation thresholds of two-dimensional continuum systems of rectangles~\cite{Li2013PRE} and ellipses~\cite{Li2016PhysA} for a wide range of aspect ratios from $\varepsilon = 1$  to $\varepsilon = 1000$  have been reported. Both ellipses and rectangles transform into sticks when $\varepsilon = \infty$. When $\varepsilon = 1$, a rectangle is simply a square, while an ellipse is a disc. Currently, the best known value of the percolation threshold of zero-width sticks of equal length that are randomly oriented and placed onto a plane, is $n_c^\times =5.637\,285\,8(6)$~\cite{Mertens2012PRE}.  By convention, the value of $a$ for sticks is taken as equal to $l^2$, where $l$ is the length of the stick.  The best known value of the percolation threshold of discs, i.e., ellipses with $\varepsilon =1$, is $\eta_c^\circ = 1.128\,087\,37(6)$, respectively $n_c^\circ = \eta_c^\circ /\left( \pi r^2 \right) = 1.436\,345\,25(8)$~\cite{Mertens2012PRE}.
A calculation has been presented for the excluded area between penetrable rectangles in 2D as a function of the aspect ratio and orientational order parameter~\cite{Chatterjee2015JSP}. The percolation threshold was found to rise with increases in the degree of particle alignment.
For isotropically distributed systems, the percolation thresholds for different values of the aspect ratio are in close agreement with findings from Monte Carlo simulations~\cite{Li2013PRE}.
Recently, percolation thresholds of superellipses have been reported~\cite{Lin201917PT}.  In a Cartesian coordinate system, the equation of a superellipse is
\begin{equation}\label{eq:supereelips}
  \frac{|x|^{2m}}{a^{2m}} + \frac{|y|^{2m}}{b^{2m}} = 1,
\end{equation}
where $a$ and $b$ are the semimajor lengths in the direction of  the $x$ and $y$ axes and $m$ is the shape parameter. $m=1$ corresponds to an ellipse while $m=+\infty$ corresponds to a rectangle. Percolation thresholds as the total fractions of the plane covered by the particles, $\phi_c$, have been presented for 14 shapes, for each of 6 aspect ratios~\cite{Lin201917PT}.

Although the percolation of spherocylinders has been studied~\cite{Xu2016PRE}, to the best of our knowledge, the percolation thresholds  for their two-dimensional analogs, i.e., discorectangles, have not yet been presented in the literature. The goal of the present work was to obtain the dependencies of the percolation thresholds of randomly placed and oriented discorectangles on their aspect ratios. The rest of the paper is constructed as follows. In Section~\ref{sec:methods}, the technical details of the simulations and calculations are described. Section~\ref{sec:results} presents our main findings. Section~\ref{sec:concl} summarizes the main results.

\section{Methods}\label{sec:methods}

We used the union--find algorithm~\cite{Newman2000PRL,Newman2001PRE} to check for any occurrences of wrapping clusters. In our study,  we used the version of the union--find algorithm adapted for continuous percolation~\cite{Li2009PRE,Mertens2012PRE}.

\begin{figure}[!htb]
  \centering
  \includegraphics[width=0.75\columnwidth]{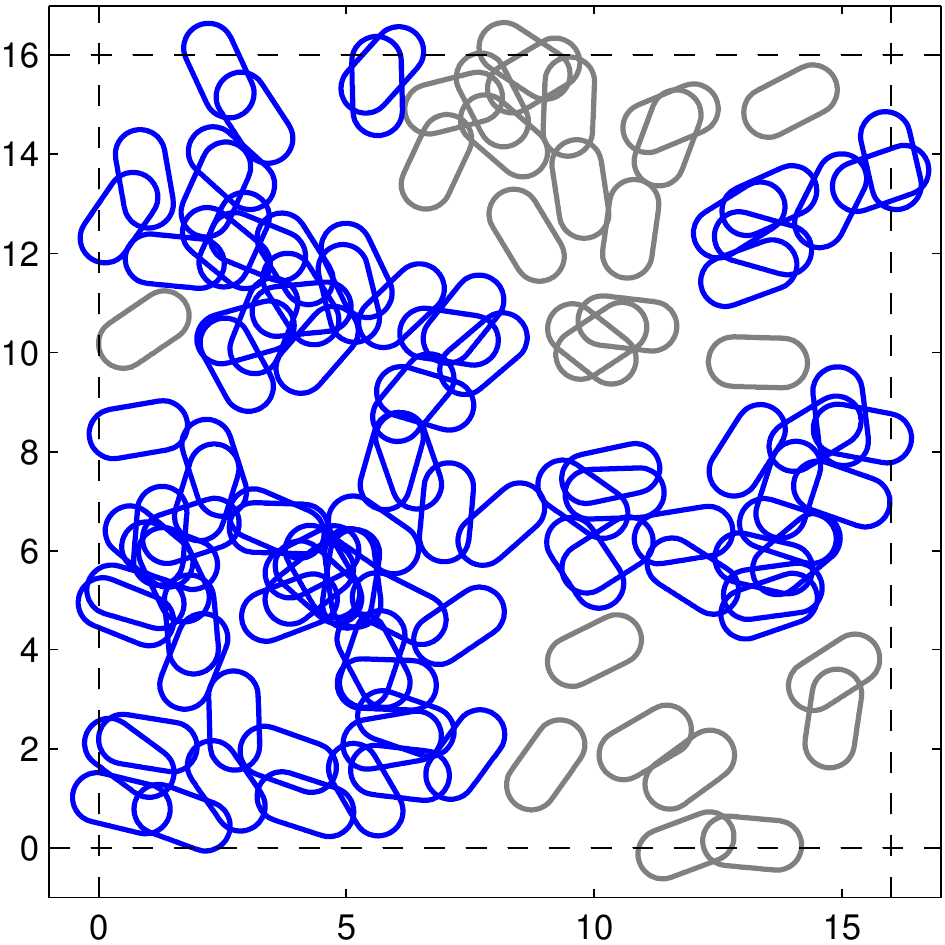}
  \caption{Example of a system of discorectangles ($\varepsilon = 2$) exactly at the percolation threshold. The linear system size is $16l$. The incipient wrapping cluster is highlighted.\label{fig:cluster}}
\end{figure}

Discorectangles with $l =1$ were added one by one randomly, uniformly, and isotropically onto a  substrate of size $L \times L$ having periodic boundary conditions (PBCs), i.e., onto a torus, until a  cluster wrapping around the torus in two directions had arisen. In this case, the desired number density, $n$, is
\begin{equation}\label{eq:numdens}
  n = \frac{N}{L^2}.
\end{equation}
Intersections of the discorectangles were allowed (figure~\ref{fig:cluster}). For each given system size, $L$, and number of deposited discorectangles, $N$,  $10^5$ independent runs were performed to obtain the probability  of percolation, $R^{(c)}_{N,L}$. Here, the superscript $c$ means a used criterion, viz., $h, v$, or $b$ mean that the cluster winds the torus in the horizontal, or vertical direction, or in both directions, respectively.

To obtain the probability $R^{(c)}(\eta,L)$ of percolation in the grand canonical ensemble, we convolved $R^{(c)}_{N,L}$ with the Poisson distribution~\cite{Li2009PRE,Mertens2012PRE}.
\begin{equation}\label{eq:convolution}
  R^{(c)}(\eta,L)= \sum_{N=0}^\infty \frac{\lambda^N \mathrm{e}^{-\lambda}}{N!} R^{(c)}_{N,L}.
\end{equation}
The weights in Eq.~\eqref{eq:convolution}
$w_N (\lambda)={\lambda^N}/{N!}$
can be calculated using the recurrent relations~\cite{Mertens2012PRE},
\begin{equation}\label{eq:weight1}
  w_{\bar{N}-k} =
  \begin{cases}
    1, & \mbox{for } k=0, \\
    \frac{\bar{N}- k + 1}{\lambda} w_{\bar{N} -k +1}, & \mbox{for } k=1,2,\dots,
  \end{cases}
\end{equation}
and
\begin{equation}\label{eq:weight2}
  w_{\bar{N}+k} =
  \begin{cases}
    1, & \mbox{for } k=0, \\
    \frac{\lambda}{\bar{N}+ k} w_{\bar{N} + k - 1}, & \mbox{for } k=1,2,\dots,
  \end{cases}
\end{equation}
herewith the relation
$$
\sum_{N=0}^\infty \frac{\lambda^N }{N!} =  \sum_{N=0}^\infty w_N (\lambda) = \mathrm{e}^\lambda, \quad  \forall \lambda > 0
$$
should be borne in mind. Here, $\bar{N} = \lfloor \lambda \rfloor$.
Therefore, the convolution can be calculated as
\begin{equation}\label{eq:RNL}
R^{(c)}(\eta,L)= \sum_{N=0}^\infty w^\ast_N(\lambda) R^{(c)}_{N,L},
\end{equation}
where
\begin{equation}\label{eq:wstar}
  w^\ast_N(\lambda) = \frac{w_N(\lambda)}{\sum_{N=0}^\infty w_N(\lambda)}.
\end{equation}
The factor $\mathrm{e}^{-\lambda}$ is absent in the master equation~\eqref{eq:RNL}, since
$$
\sum_{N=0}^\infty w_N (\lambda) = \mathrm{e}^\lambda \sum_{N=0}^\infty w^\ast_N (\lambda).
$$

Conformal field theory gives exact values for the wrapping probabilities at the
transition in the limit $L \to \infty$~\cite{Pinson1994JSP,Newman2000PRL,Newman2001PRE}.
\begin{equation}\label{eq:R}
  R^{(c)}_\infty =
  \begin{cases}
    0, & \text{if } \eta < \eta_c, \\
    R^{\ast}, & \text{if } \eta = \eta_c, \\
    1, & \text{if } \eta > \eta_c,
  \end{cases}
\end{equation}
where $R^{\ast} = 0.521\, 058\, 290\dots$ is the probability of wrapping horizontally around the system and
$R^{\ast} = 0.351\, 642\, 855\dots$ is the probability of wrapping around both directions simultaneously. More precise values of $R^\ast$ including other possible criteria are presented in Ref.~\cite{Mertens2012PRE}.
This theory provides the most effective method for estimating the percolation threshold~\cite{Newman2000PRL,Newman2001PRE,Li2009PRE,Mertens2012PRE} since
\begin{equation}\label{eq:sqaling}
  \eta_c(\infty) - \eta_c(L) \propto L^{-2-1/\nu}, \text{ where } \nu= 4/3.
\end{equation}
Typically, we used systems of sizes $L = 8, 16, 32, 64$ to perform the scaling analysis. The number of independent runs was $10^5$. All results presented in Section~\ref{sec:results} correspond to the thermodynamic limit. 

\begin{figure}[!htb]
  \centering
  \includegraphics[width=\columnwidth]{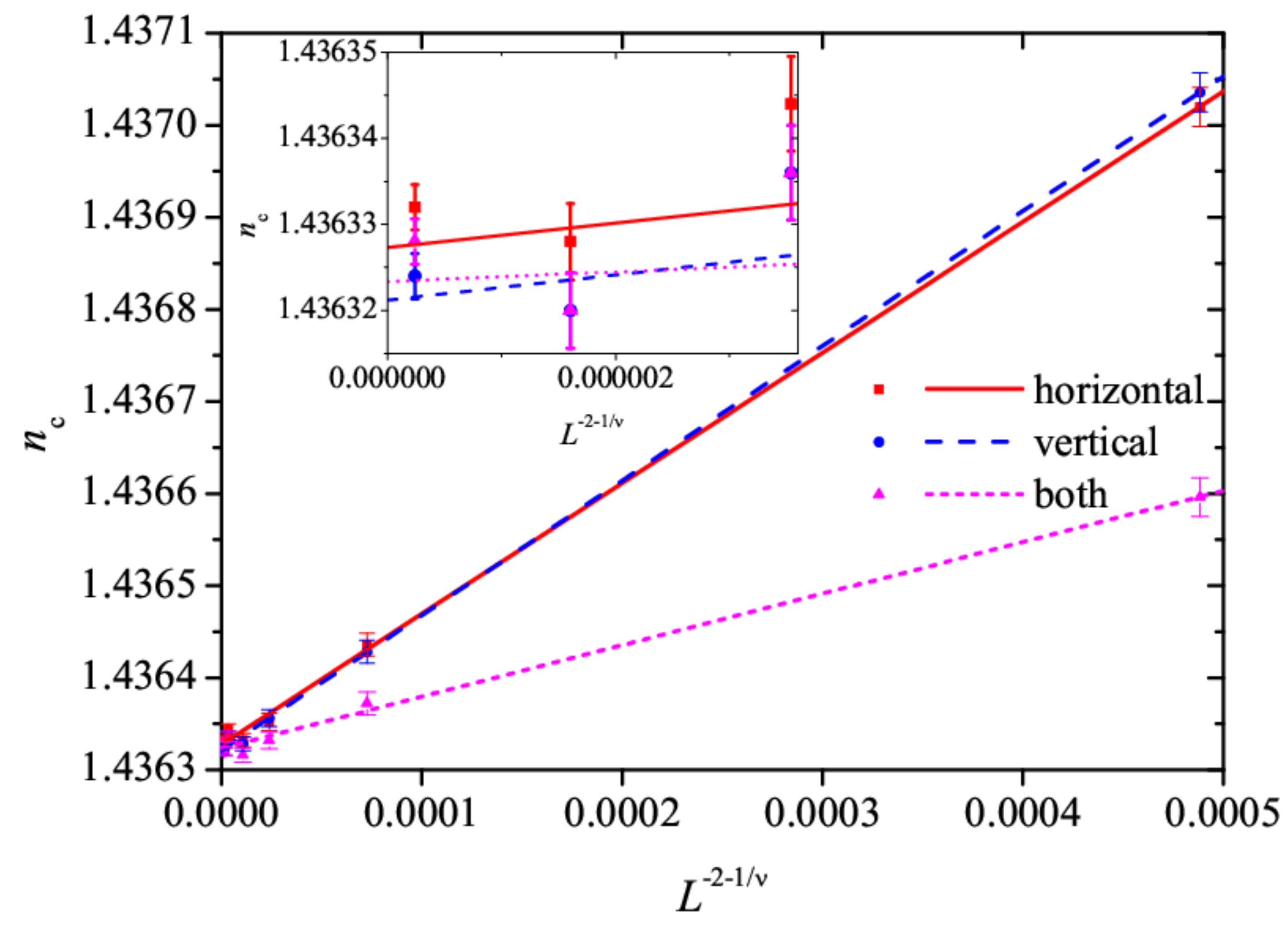}
  \caption{Example of scaling for discs, i.e., discorectangles with $\varepsilon = 1$.\label{fig:scalingdiscs}}
\end{figure}

To verify our program, we performed more accurate estimations for one particular case, viz., $\varepsilon = 1$ (discs of $r=1$). For this particular case we used $L = 8, 16, 32, 64, 128, 256$ while the number of independent runs was $10^8$ (figure~\ref{fig:scalingdiscs}).
\begin{figure}[!htb]
  \centering
  \includegraphics[width=\columnwidth]{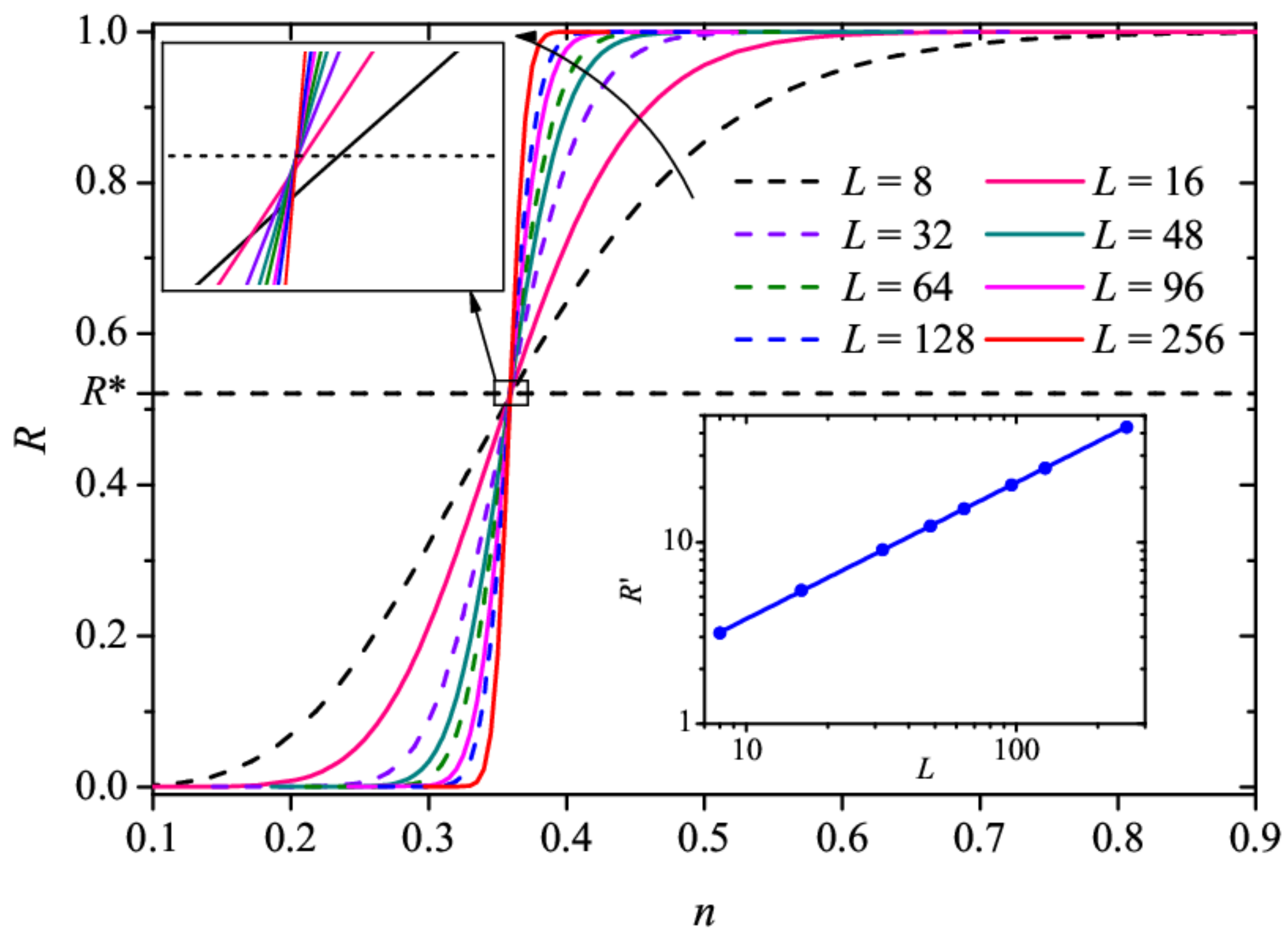}
  \caption{$R(n)$ for different system sizes. Curved arrow indicates increasing of $L$. Inset: Example of the dependency of $R'(L)$ for wrapping in the horizontal direction in log-log scale.\label{fig:slope}}
\end{figure}

Figure~\ref{fig:scalingdiscs} demonstrates an example of scaling for $\varepsilon=1$. According to Ref.~\cite{Mertens2012PRE}, the standard deviation was taken as $\approx N_{ir}^{-1/2}L^{-3/4}$, where $N_{ir}$ is the number of independent runs.  Our estimations gave $n_c^\circ = 1.436\,320(4)$ with the adjusted $R^2 = 0.99$. This estimation is reasonable close to the published values for discs of unit diameter $n_c^\circ = 1.436\,323(3)$~\cite{Li2016PhysA} and $n_c^\circ = 1.436\,345\,25(8)$~\cite{Mertens2012PRE}.

Additionally, we checked the derivative of $R$ at the percolation threshold. An example of the dependency  of $R'(L)$ for wrapping in the horizontal direction is presented in figure~\ref{fig:slope} (inset) in log-log scale. The slope is $0.752 \pm 0.002$ and this corresponds to the value of the critical exponent $\nu$.

\section{Results}\label{sec:results}

Our results are presented in Table~\ref{tab:percolationthreshold} which compares the percolation thresholds for discorectangles (our results) with known values for rectangles $n^{r}_c$ \cite{Li2013PRE} and ellipses $n^{e}_c$ \cite{Li2016PhysA} for different values of the aspect ratio.
\begin{table}[htb]
\caption{Comparison of the percolation thresholds of rectangles $n^{r}_c$ \cite{Li2013PRE}, ellipses $n^{e}_c$ \cite{Li2016PhysA}, and discorectangles (our results) for different values of the aspect ratio. Case $\varepsilon = \infty$ corresponds to the percolation of sticks~\cite{Mertens2012PRE}. The values are rounded to significant figures.\label{tab:percolationthreshold}}
\begin{ruledtabular}
\begin{tabular}{rccc}
$\varepsilon $ & $n^{r}_c$ & $n^{dr}_c$ & $n^{e}_c$\\
  \hline\\
1    & 0.982278 & 1.436 & 1.436323\\
1.5  & 1.425745 & 1.894 & 2.059081\\
2    & 1.786294 & 2.245 & 2.523560\\
3    & 2.333491 & 2.760 & 3.157339\\
4    & 2.731318 & 3.123 & 3.569706\\
5    & 3.036130 & 3.396 & 3.861262\\
6    & 3.278680 & 3.612 & 4.079359\\
7    & 3.477211 & 3.787 & 4.249158\\
8    & 3.643137 & 3.933 & 4.385303\\
9    & 3.784321 & 4.057 & 4.497044\\
10   & 3.906022 & 4.163 & 4.590416\\
15   & 4.329848 & 4.530 & 4.894745\\
20   & 4.584535 & 4.749 & 5.062313\\
30   & 4.878091 & 5.000 & 5.241522\\
50   & 5.149008 & 5.229 & 5.393863\\
100  & 5.378856 & 5.422 & 5.513464\\
200  & 5.504099 &       & 5.612260\\
1000 & 5.609947 &       & 5.624756\\
  \hline
$\infty$ & & 5.6372858 &\\
\end{tabular}
\end{ruledtabular}
\end{table}

\begin{figure}[!htb]
  \centering
  \includegraphics[width=\columnwidth]{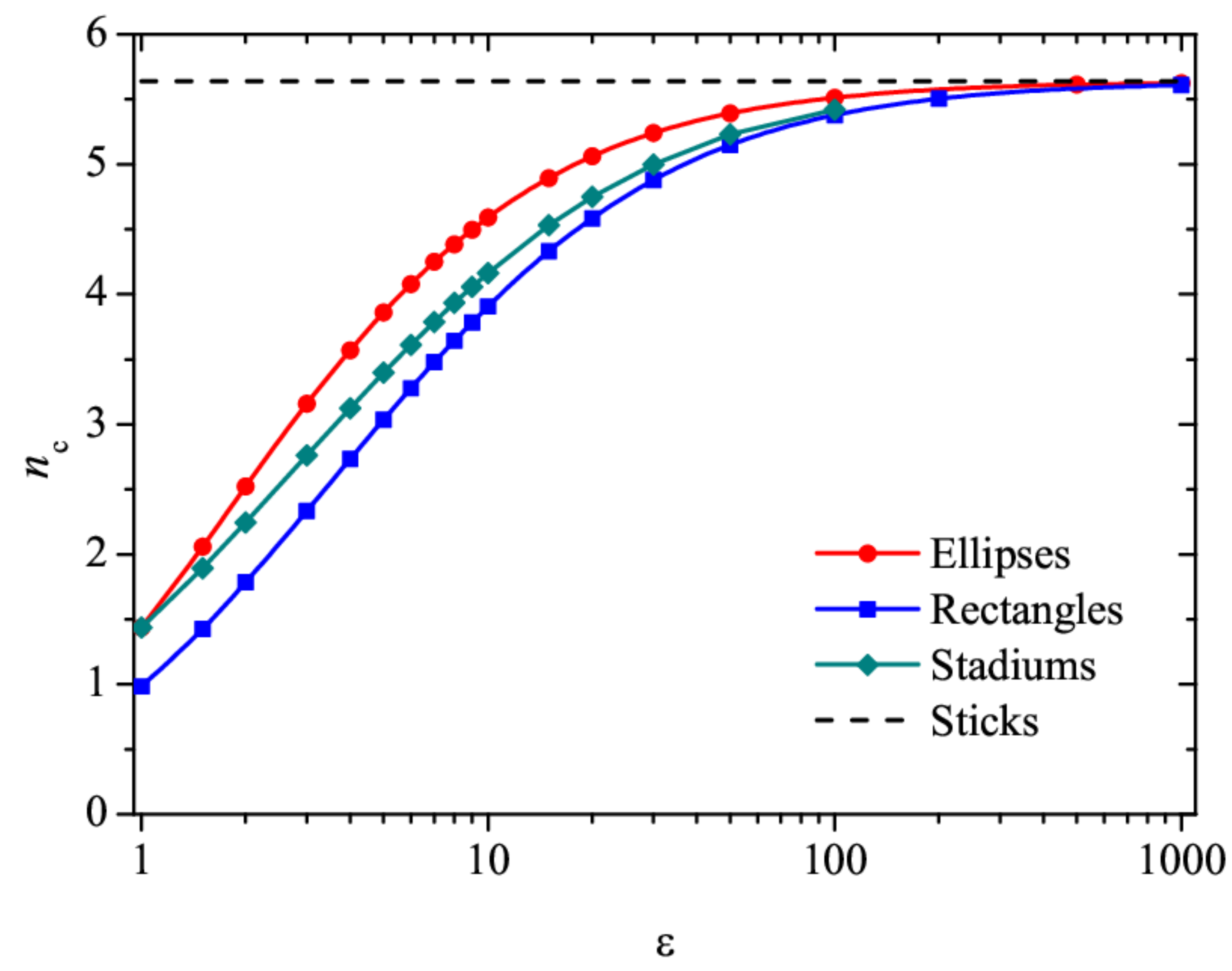}
  \caption{\label{fig:percolationthreshold}Dependencies of the percolation threshold, $n_c$, on the aspect ratio, $\varepsilon$, for rectangles~\cite{Li2013PRE}, ellipses~\cite{Li2016PhysA}, and discorectangles (our results) in semi-log plot with a logarithmic scale on the $\varepsilon$ axis, and a linear scale on the $n_c$ axis. The horizontal dashed line corresponds to the percolation of sticks ($\varepsilon = \infty$)~\cite{Mertens2012PRE}. The error bars are of the order of the marker size when not shown explicitly.}
\end{figure}

Figure~\ref{fig:percolationthreshold} demonstrates the dependencies of the percolation threshold, $n^{dr}_c$, on the aspect ratios of discorectangles, $\varepsilon$. The dependencies for rectangles $n^{r}_c(\varepsilon)$ \cite{Li2013PRE} and ellipses $n^{e}_c(\varepsilon)$ \cite{Li2016PhysA} are shown for comparison. For any value of $\varepsilon$, the critical number density increases as the aspect ratio increases. When $\varepsilon =1$, the discorectangle is simply a disc, hence, the percolation threshold of such discorectangles equals the percolation threshold of discs. When $\varepsilon \to \infty$, ellipses, rectangles, and discorectangles all tend to sticks. Thus their percolation thresholds approach the percolation threshold of zero-width sticks. For any values of $\varepsilon$, the percolation threshold of discorectangles is situated between the percolation thresholds of ellipses (upper boundary) and rectangles (lower boundary).

\section{Conclusion}\label{sec:concl}
By means of computer simulation and scaling analysis, we studied the percolation of  discorectangles on a torus. The dependencies of the percolation threshold, $n^{dr}_c$, on the aspect ratio, $\varepsilon$, have been obtained in the thermodynamic limit. Comparison with known results for rectangles~\cite{Li2013PRE}, $n^{r}_c$, and ellipses~\cite{Li2016PhysA}, $n^{e}_c$, evidenced that $$ n^{r}_c(\varepsilon) < n^{dr}_c(\varepsilon) \leqslant n^{e}_c(\varepsilon).$$
Naturally,  $n^{dr}_c(0) = n^{e}_c(0)$ since, in this case, each of these  shapes is simply a disc. The value of $ n^{dr}_c(\varepsilon)$ tends to the value $n_c$ for zero-width sticks~\cite{Mertens2012PRE} when $\varepsilon \to \infty$.  Improvements to the accuracy of the obtained values of the percolation threshold will require additional time and computational resources.

\begin{figure}[!htb]
  \centering
  \includegraphics[width=0.9\linewidth]{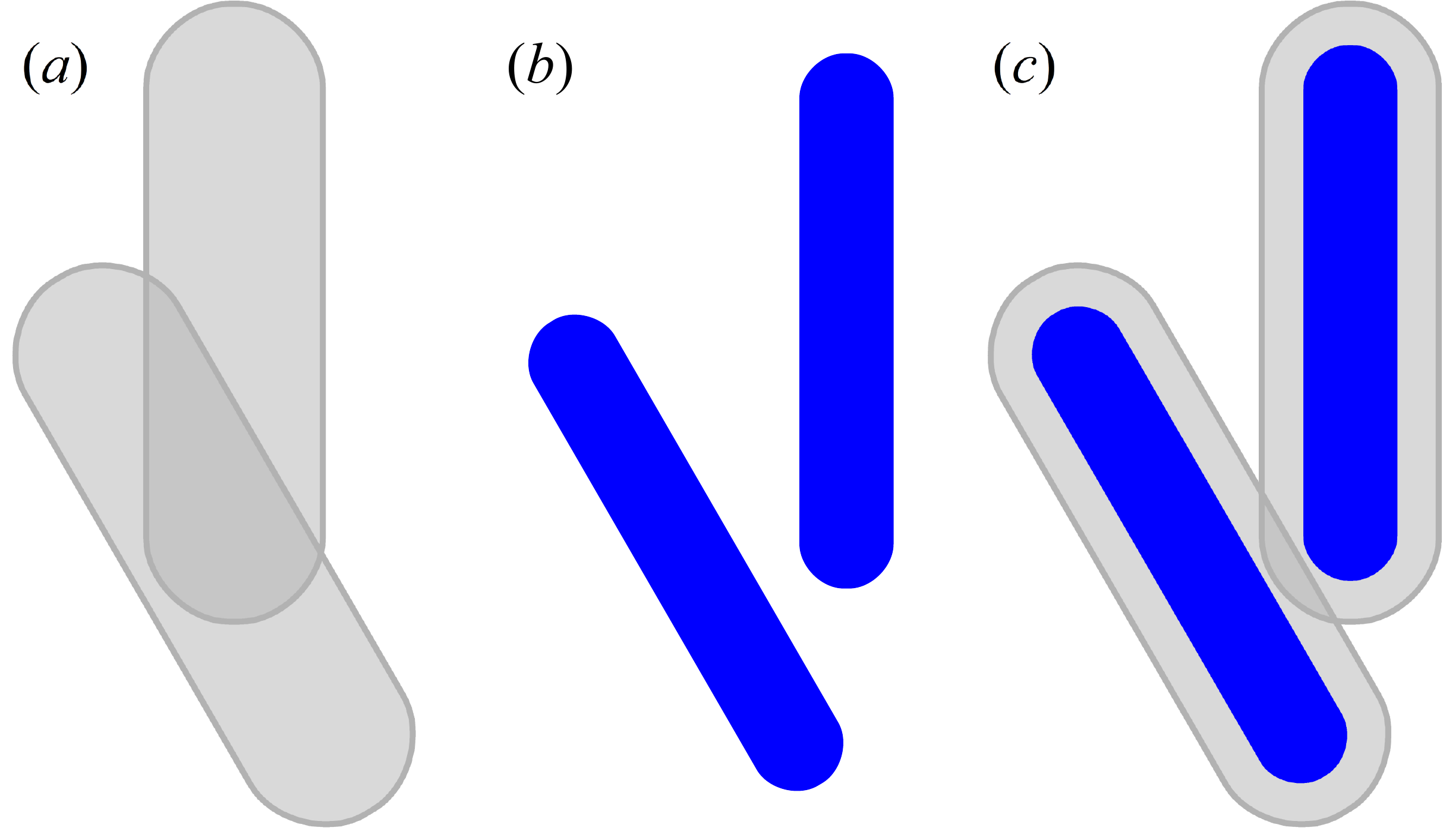}
  \caption{Permeable (a) and hard--core--soft--shell (c) particles can form a cluster while impermeable particles (b) cannot.}\label{fig:clusters}
\end{figure}

Our consideration deals with only one particular case when overlapping of particles is allowed. Such the particles are treated as permeable, overlapped particles form a cluster (Fig.~\ref{fig:clusters}a). However, other possibilities are also feasible. For instance, in random sequential adsorption~\cite{Evans1993RMP}, particles are impermeable and none cluster can occur (Fig.~\ref{fig:clusters}b).

An intermediate possibility is so-called connectedness percolation of non-overlapping particles~\cite{Otten2011JCP,Drwenski2018JCP}. Two non-overlapping particles are assumed to be connected when the shortest distance between them does not a exceed a certain value, i.e., so-called cutoff distance~\cite{Lee1988JCP}. This case can be also treated as a hard--core--soft--shell model (Fig.~\ref{fig:clusters}c). Naturally, in this case, the percolation threshold have to significantly depend on the cutoff distance. 

Both length dispersity and alignment of particles may affect the percolation threshold~\cite{Otten2009PRL,Otten2011JCP,Tarasevich2018PREsticks}. In the case of permeable discorectangles, these effects needs an additional examination.

\begin{acknowledgments}
We acknowledge the funding from the Ministry of Science and Higher Education of the Russian Federation, Project No.~3.959.2017/4.6.
\end{acknowledgments}

\bibliography{dispersity}

\end{document}